\documentclass{iopart}
\usepackage{iopams}  
\usepackage{a4wide}
\usepackage{graphicx,subfigure}

\newcommand{\be}{\begin{equation}}
\newcommand{\ee}{\end{equation}}
\newcommand{\bea}{\begin{eqnarray}}
\newcommand{\eea}{\end{eqnarray}}

\newcommand{\bel}[1]{\begin{equation}\label{#1}}
\newcommand{\beal}[1]{\begin{eqnarray}\label{#1}}

\begin{document}

\title{Phase portrait of a matter bounce in Ho\v{r}ava-Lifshitz cosmology}
\author{E.~Czuchry}
\address{
$^1$Instytut
Problem\'ow J\c{a}drowych, ul. Ho\.za 69, 00-681 Warszawa, Poland}
\ead{eczuchry@fuw.edu.pl}

\begin{abstract}

The occurrence of a bounce in FRW cosmology requires modifications of general
relativity. An example of such a modification is the recently proposed
Ho\v{r}ava-Lifshitz theory of gravity, which includes a ``dark radiation'' term
with a negative coefficient in the analog of the Friedmann equation. This letter
describes a phase space analysis of
models of this sort with the aim of determining to what extent bouncing
solutions can occur.  It is found that they are is possible, but not generic
in models under consideration. Apart from previously known bouncing solutions
some new ones are also described. Other interesting solutions found include ones
which describe a novel sort of quasi stationary, oscillating universes.
\end{abstract}

\maketitle

\section{Introduction}

The standard $\Lambda$CDM model has solved many issues in cosmology. However,
in spite of all this success, it also leaves a number of issues unaddressed.
Perhaps the most significant is the problem of initial singularity, where
general relativity breaks down. There have been many attempts to modify
Einstein's theory to avoid this singularity. Some are made at classical level,
some involve quantum effects. Examples include the ekpyrotic/cyclic model
(\cite{Ekp,Cyclic,pyrotech}) and loop quantum cosmology (\cite{ap1,ap2,ap3}), which
replace the Big Bang with a Big Bounce. Attempts to address these issues at
the classical level include braneworld scenarios (\cite{bw,rs}), where the
universe goes from an era of accelerated collapse to an expanding era without
any divergences or singular behavior.  There are also higher order
gravitational theories and theories with scalar fields (see \cite{bc} for a
review of bouncing cosmologies). It is however fair to say that the issue of the
initial singularity remains one of the key questions of early Universe
cosmology, and the idea that it is avoided due to a bounce is remains an elusive
(and promising) notion. As discussed below, it is clear that close to the
singularity the Friedmann equation has to be modified for a bounce to be
possible.

In recent months much effort has been devoted to studies of a proposal for a UV
complete theory of gravity due to Ho\v{r}ava~\cite{hor2,hor1,horava,Horava:2009if}.  Because in the
UV the theory possesses a fixed point with an anisotropic, Lifshitz scaling
between time and space, this theory is referred to as Ho\v{r}ava-Lifshitz
gravity.  Soon after this theory was proposed many specific solutions of this
theory have been found, including cosmological ones
(\cite{hc1, hc2, Saridakis:2009bv, Nastase:2009nk, Mukohyama:2009zs, Minamitsuji:2009ii, Wang:2009rw,Takahashi:2009wc}).  It was also realized that the analog of the
Friedmann equation in HL gravity contains a term which scales in the same way as
dark radiation in braneworld scenarios \cite {hc1,hc2,Saridakis:2009bv} and gives a negative
contribution to the energy density. Thus, at least in principle it is possible
to obtain non-singular cosmological evolution within Ho\v{r}ava theory, as it
was pointed out in \cite{hc1,Saridakis:2009bv,Brandenberger:2009yt}.

Although there is presently much discussion of possible problems and
instabilities of Ho\v{r}ava-Lifshitz gravity
\cite{Charmousis:2009tc, LiPang, Sotiriou:2009bx, Bogdanos:2009uj}, it is still very interesting to
perform a detailed investigation of the influence of the additional terms in the
Friedmann equation in HL gravity on the existence and stability of a
cosmological bounce. In this paper we are going to analyze how these terms
affect the dynamics of the system using phase portrait techniques described in
\cite{frolov, frolov2}, and then compare the results with those valid in
standard cosmology. For purpose of illustration we 
will assume that matter in the pre-bounce epoch is described
by a scalar field $\varphi$ with a quadratic potential.
  In order to concentrate on modifications created by the
``dark radiation'' terms, we will set cosmological constant $\Lambda=0$. Such scenario may be also considered  as an approximation to a general case with non-vanishing $\Lambda$, valid
in the regime of small scale factor $a$, when standard curvature and cosmological constant terms (vanishing for  $\Lambda=0$) become negligible. Thus,
the present analysis can be regarded as an exploration of the cosmologies with
modified equations of motion, where the particular modifications considered are
inspired by Ho\v{r}ava cosmology. In particular, the main question addressed is
the impact of these modifications on the existence of a bounce, which is
otherwise not possible.

Related analysis of Ho\v{r}ava-Lifshitz cosmology have recently appeared in
\cite{ps1} and \cite{ps2}, which we become aware of while this work was being
typed. Those papers address a somewhat different set of issues from what we have
pursued.  The analysis presented in \cite{ps1} and \cite{ps2} consider the full
4-dimensional phase space of HL cosmology. The results presented here focus on
the region close to where the scale factor vanishes, which admits a critical
simplification: the number of dynamical equations under study can be reduced
from 4 to 3 (as discussed in more detail in Section 3). This makes it possible
to visualize the possible phase space trajectories in a 3-dimensional space.

The structure of this note is following: in Section 2. we briefly sketch
Ho\v{r}ava-Lifshitz gravity and cosmology. In Section 3. the possibility of
bounce is discussed. In Section 4 we discuss phase portraits of the discussed
system of equations and describe different families of phase trajectories.

\section{Ho\v{r}ava-Lifshitz cosmology}

The metric of Ho\v{r}ava-Lifshitz theory, due to  anisotropy in UV,  is written
in the $(3+1)$-dimensional ADM formalism:
\be
\label{eq1} ds^2=-N^2 dt^2 +g_{ij}(dx^i-N^idt)(dx^j-N^jdt),
\ee
where $N$, $N_i$ and $g_{ij}$ are dynamical variables.
The action of Ho\v{r}ava-Lifshitz theory is~\cite{horava}
\bea
\label{eq3}
I &=& \int dtd^3x ({\cal L}_0 +{\cal L}_1), \label{action}\\
 {\cal L}_0 &=& \sqrt{g}N \left \{\frac{2}{\kappa^2}
(K_{ij}K^{ij}-\lambda K^2) +\frac{\kappa^2\mu^2 (\Lambda
R-3\Lambda^2)}{8(1-3\lambda)}\right \},  \nonumber \\
 {\cal L}_1  &=& \sqrt{g}N \left \{\frac{\kappa^2\mu^2(1-4\lambda)}{32(1-3\lambda)}R^2
-\frac{\kappa^2}{2\omega^4}Z_{ij} Z^{ij} \right\} ,\nonumber
\eea
where $K_{ij}=\frac1N \left[\frac12\dot g_{ij}-\nabla_{(i}N_{j)}\right]$ is extrinsic
curvature of  a spacelike hypersurface with a fixed time, a dot denotes a
derivative with respect to $t$ and covariant
derivatives defined with respect to the spatial metric $g_{ij}$. Moreover
\be
Z_{ij}=C_{ij}-\frac{\mu\omega^2}{2}R_{ij} .
\ee
$\kappa^2$, $\lambda$, $\mu$, $\omega$ and $\Lambda$ are
constant parameters and the Cotton tensor, $C_{ij}$, is defined by
\be
\label{eq4} C^{ij}=\epsilon^{ikl} \nabla_k \left (R^j_{\
l}-\frac{1}{4}R\delta^j_l\right) = \epsilon^{ikl}\nabla_k R^j_{\ l}
-\frac{1}{4}\epsilon^{ikj}\partial_kR.
\ee
In (\ref{eq3}),  ${\cal L}_0$ is the kinetic part of the action, while ${\cal
  L}_1 $ gives the potential of the theory in the so-called
``detailed-balance" form.\\

Matter may be added by introducing a scalar field $\varphi$ (\cite{hc1,hc2}) with energy density $\rho$ and pressure $p$. The action for matter is
\be
I_m = \int dtd^3x \sqrt{g}N{\cal L}_m.\label{lm}
\ee
The matter Lagrangian ${\cal L}_m$ depends on the scalar
matter field $\varphi$ and the 4-dimensional metric.  ${\cal L}_m$:
\be
{\cal L}_m=\frac{3\lambda-1}2\left( \frac1{2N^2} (\dot{\varphi}^2-N^i\partial_i\varphi) -  V(\varphi)\right)
\ee
{This allows to define the energy density and pressure of the scalar field in the following way:
\bea
\rho&=&\frac{3\lambda-1}4\dot{\varphi}^2+V(\varphi),\\
p&=&\frac{3\lambda-1}4\dot{\varphi}^2-V(\varphi)
\eea
In numerical calculations presented further on a specific form of the scalar potential will be assumed (see eqn. (\ref{scalar})). 

Comparing the action  of Ho\v{r}ava-Lifshitz theory to the Einstein-Hilbert
action of general relativity, one can see that the speed of light, Newton's
constant and the cosmological constant
are
\be\label{eq5}
c=\frac{\kappa^2\mu}{4}\sqrt{\frac{\Lambda}{1-3\lambda}}, \ \
G=\frac{\kappa^2 c}{32\pi}, \ \  \Lambda_E=-\frac{3\kappa^4\mu^2}{3\lambda-1}\frac{\Lambda^2}{32},
\ee
respectively. Setting dynamical constant $\lambda=1$, reduces the first three
terms in (\ref{eq3})  to the usual ones of
Einstein's relativity and matter Langrangian in (\ref{lm}) to the usual scalar field  action in curved
space-time.

The equations for Ho\v{r}ava-Lifshitz cosmology are obtained by  imposing condition of
homogeneity and isotropy of the metric. Precisely,
the equations of motion are obtained by varying the action (\ref{action}) with
respect to $N$, $a$, and $\varphi$, and setting $N = 1$ at the end of the
calculation, leading to
\bea
H^2 &=&\frac{\kappa^2 \rho}{6(3\lambda-1)} +\frac{\kappa^4\mu^2\Lambda}{8(3\lambda-1)^2} \frac{ k}{a^2} - \frac{\kappa^4\mu^2 }{16(3\lambda-1)^2}\left(\Lambda^2+\frac{ { k}^2 }{a^4}\right)
, \label{hc1} \\
{\dot H}  &=& -\frac{\kappa^2(\rho+p)}{4(3\lambda-1)} -\frac{\kappa^4\mu^2\Lambda}{8(3\lambda-1)^2} \frac{ k}{a^2} + \frac{\kappa^4\mu^2 }{32(3\lambda-1)^2} \frac{ { k}^2 }{a^4}, \label{hc2}
\eea
and also equation of motion for the scalar field:
\be
{\ddot \varphi} + 3 H {\dot \varphi} + \frac2{3\lambda-1}V^{\prime}  =  0  , \label{mattereq}
\ee
 where $H = {\dot a}/a$, a prime denotes the derivative
with respect to scalar field $\varphi$. 
The significant new terms in the above equations of motion are the
$(1/a^4)$-terms on the right-hand sides of (\ref{hc1}) and
(\ref{hc2}). They are reminiscent of the  dark radiation term in braneworld cosmology \cite{BDEL}
and are present only if the spatial curvature of the metric is non-vanishing.

Equations  (\ref{hc1}) and
(\ref{hc2}) show different behavior for different ranges of the $\lambda$-parameter:  $\lambda>1/3$, $\lambda=1/3$ and $\lambda<1/3$.  It was shown in \cite{bs} that $1>\lambda>1/3$ leads to ghost  instabilities in the IR limit of the theory. Solution to this problem proposed in \cite{lmp} results in instabilities re-emerging at UV region.  However, this range of $\lambda$ is exactly the flow-interval between the UV and IR regimes. Thus 
the only physically interesting case that remains,
allowing for a possible flow towards GR -- at $\lambda=1$ --  is the regime $\lambda\ge1$. Region $\lambda\le 1/3$ is disconnected from $\lambda=1$ and cannot be included in realistic considerations.

Thus we will remain in the phenomenologically range relevant  $\infty>\lambda\ge1$. In this case the value of $\lambda$ -- a dimensionless coupling of the theory -- may be included in rescalling of the parameter $\kappa$. In general, $\lambda$ runs -- logarithmically in the UV -- and may eventually reach one the three IR fixed points (\cite{hor1}): $\lambda=1/3$, $\lambda=1$ or $\lambda=\infty$, the first one excluded by the existence of instabilities.

\section{Existence of bounce}

New terms in the cosmological equations introduce the possibility of a bounce. The
form of (\ref{hc1}), with $k=\pm1$ implies that it is possible that $H=0$ at
some moment of time.
This is a necessary condition for the realization of the bounce. It was pointed
out in \cite{hc1}, that it may happen in the presence of matter, at the critical
time $t_*$, $a=a_*$, when the critical energy density is equal to
\be
\rho=\rho_*= \frac{3\kappa^2\mu^2}{3\lambda-1}\left(-\frac{\Lambda}{4}\frac{k}{a_*^2}+ \frac{\Lambda^2}{8}
+\frac{1}{8}\frac{ { k}^2}{ a_*^4}
\right),
\ee
which is determined by the couplings of the theory.

From the continuity
equation it follows that at the bounce point $\dot H
> 0$. Therefore a transition from a contracting to
an expanding phase may be possible. It was shown in \cite{Brandenberger:2009yt}
that the necessary condition for a cosmological
bounce is that  the energy density of
regular matter increases less fast than $a^{-4}$ as the scale
factor decreases and $(\frac{\rho}{12} - p)  >  0$ .

We begin our considerations during a  contracting phase. At the beginning the
scale factor is quite large and the contribution of dark radiation to the total
energy density is quite small. As the universe contracts, the energy density
increases and the scale factor decreases rapidly. When a critical density is
achieved, a big bounce is about to take place.

 One would expect that near the bounce the leading term in (\ref{hc1}) and (\ref{hc2})
 would be the dark
radiation one,  with  curvature and cosmological constant terms neglectable small. 
Actually the latter terms generally vanish when HL cosmological constant $\Lambda= 0$. 
In case of non-vanishing $\Lambda$ these two terms may be neglected when the scale factor  $a$ is sufficiently small.  
  Specifically, assuming for a moment an equation of state of the form $p = w
  \rho$ with constant $w$, it is well known that $H^2$, $\dot{H}$ and $\rho$
  scale as $a^{-3(1+w)}$. Therefore we may keep the density term and omit the
  curvature term $\sim1/a^2$ if $w>-\frac13$.  In the case of a quadratic
  potential considered below (for which $w\neq \textrm{const}$, so the above
  argument does not directly apply) we have checked numerically that in all
  bounce scenarios discussed in this paper, this approximation is valid near
  bounce point (up to $10^{-7}$).

We will model the matter sector in this pre-bounce
epoch by assuming it is described
by a scalar field $\varphi$ with a potential
\be
V(\varphi) =  \frac{1}{2} m^2 \varphi^2 \label{scalar} .
\ee
For calculational simplicity we put $m=1$.

This way and inserting $\alpha= 2/\kappa^2$, we have the following equations modeling  bounce in the
Ho\v{r}ava-Lifshitz cosmology. \bea {\dot H}  &=& -
\frac{\kappa^2}{4(3\lambda-1)}\dot{\varphi}^2+\frac{\kappa^4\mu^2}{8(3\lambda-1)^2}
\frac{k^2}{a^4}, \label{hc11}\\\
H^2  &=&  \frac{\kappa^2}{12(3\lambda-1)}(\dot{\varphi}^2+\varphi^2)-\frac{\kappa^4\mu^2}{16(3\lambda-1)^2}
\frac{k^2}{a^4}. \label{hc22}
\eea
The value of $\kappa^2$  may be expressed in terms of cosmological constants
(\ref{eq5}), we will work in units such that $8\pi G=1$ and $c=1$. Then
\be
\kappa^2=32\pi G c,
\ee
and the values of $\mu$ are left arbitrary.  Therefore the Friedmann equations take  the following
form near the bounce:
\bea
{\dot H}  &=& -\frac1{3\lambda-1}\dot{\varphi}^2+\frac{2\mu^2 k^2}{(3\lambda-1)^2a^4} \label{or1}\\
H^2 &=&  \frac1{3(3\lambda-1)} (\dot{\varphi}^2+\varphi^2)-
\frac{\mu^2k^2}{(3\lambda-1)^2a^4},\label{or2}
\eea
Additionally,  completing dynamics of the system, there is the equation of motion for the scalar field and the definition of the Hubble parameter:
\bea
\ddot{\varphi} &=& -\frac2{3\lambda-1}\varphi- 3 \dot{\varphi} H,\label{em10} \\
\dot{a}&=& aH.\label{em20}
\eea
The value of the parameter $\mu$ may be kept arbitrary. This parameter does not alter
solutions of the system (\ref{or1}-\ref{em20}), but it specifies values of $a$ on
an obtained trajectory.

\section{Phase portrait}

\subsection{Phase space}
The local geometry of the phase portrait is characterized by the nature and
position of its critical points. These  points
are locations where the
derivatives of all the dynamic variables, i.e. the r.h.s.  of
(\ref{em1}-\ref{em3}), vanish.  Moreover, they are the only points where phase
trajectories may start, end, or intersect. They can also begin or end in
infinity, and then -- after a suitable coordinate transformation projecting the
complete phase space onto a compact region (so called Poincar\'{e} projection)
-- there may be well defined {\it infinite} critical points. The set of finite and
infinite critical points and their characteristic, given by the properties of
the Jacobian matrix of the linearized equations at those points, provides a
qualitative description of the given dynamical system.

Dynamics of our system is described by the following set of first order ODE's:
\bea
u&=&\dot{\varphi}\label{pe10}\\
\dot{u} &=& -\frac2{3\lambda-1}\varphi- 3 u H,\label{pe20} \\
\dot{a}&=& aH\label{pe30}\\
{\dot H}  &=& -\frac1{3\lambda-1}u^2+\frac{2\mu^2 k^2}{(3\lambda-1)^2a^4} \label{pe40}
\eea
plus the constraint equation:
\be
H^2 =  \frac1{3(3\lambda-1)} (u^2+\varphi^2)-
\frac{\mu^2k^2}{(3\lambda-1)^2a^4},\label{pe50}
\ee
If spatial curvature $k=0$ one may consider a 2-dimensional subsystem:
\bea
u&=&\dot{\varphi}\\
\dot{u} &=& -\frac2{3\lambda-1}\varphi- 3 u H, \\
{\dot H}  &=& -\frac1{3\lambda-1}u^2\\
\eea
with a constraint equation
\be
H^2 =  \frac1{3(3\lambda-1)} (u^2+\varphi^2)\label{flat-c}
\ee
If $k\neq0$ one may also consider a subsystem on variables $(\varphi,u,H)$, obtained via reduction of the original system with respect to constraint (\ref{pe50}). Namely, substituting
\be
\frac{ \mu^2k^2}{(3\lambda-1)^2a^4}=\frac1{3(3\lambda-1)} (u^2+\varphi^2)-H^2
\ee
into the equation for $\dot{H}$ and omitting equation on dynamics of $a$ leads the following set of equations:
\bea
u&=&\dot{\varphi}\label{em1}\\
\dot{u} &=& -\frac2{3\lambda-1}\varphi- 3 u H,\label{em2}\\
\dot{H}&=& \frac2{3(3\lambda-1)} (\varphi^2-\frac{u^2}2)-2H^2,\label{em3}
\eea
This is a reduced 3-dimensional subset of (\ref{pe10}-\ref{pe50}) on variables $(\varphi,u,H)$. If one wants to obtain also dynamics of $a$, he needs to add to this system equation $\dot{a}=aH$ and also the constraint equation (\ref{pe50}).

 In subsequent considerations we shall focus on a case $k\neq 0 $ (when HL corrections play a significant role) and a phase portrait of solution of the system (\ref{em1}-\ref{em3})
in space of $(\varphi,u, H)$, following similar
a similar procedure as that described in (\cite{frolov,frolov2}). Reducing dimensionality of phase space  enables 3D phase portrait visualizations.  Moreover,  we will discuss shortly also a case $k=0$, which play a role of a limiting case of $k\neq0$ dynamics.

We start by rewriting equations (\ref{em1}-\ref{em3}) in
terms of the variables
\be
x \equiv \varphi;\;y \equiv \dot{\varphi};\;z \equiv \frac{\dot {a}}{ a},
\ee
which gives three ``evolution'' equations
\bea
\dot{x} &=& y ,\label{pe1}\\
\dot{y} &=& -\frac2{3\lambda-1}x- 3 y z,\label{pe2} \\
\dot{z}&=& \frac2{3(3\lambda-1)} (x^2-\frac{y^2}2)-2z^2, \label{pe3}
\eea

The space of solution of the above dynamical system is a 3D region of the phase space $(x,y,z)$. This region is bounded by a 2D surface defined by a constraint equation (\ref{flat-c}) -- space of trajectories of a flat universe
($k=0$). This limiting surface is a double cone, with the upper branch corresponding to expansion
and lower one to contraction. Those two branches  connect at a point: (0,0,0), which is a critical
point (see below). }Hence there are no trajectories passing from one branch of
the cone to the other. For $k=\pm1$ all trajectories lie between the branches of
this cone.  This cone is also a limiting surface for trajectories 
with large $a$. The further a trajectory lies from this cone, the smaller are
the values of $a$ along it.

 Moreover, with the value of $\lambda$ varying between $1$ and $\infty$, the
  double cone reduces to the surface $z=0$ when $\lambda=\infty$. Thus for finite
  values of $\lambda$ the whole dynamics of the system is contained within the
  double cone $z^2= \frac1{3(3\lambda-1)}(x^2+y^2)$, and for $\lambda=\infty$
  all phase point lay on the surface $z=0$. In the latter case the phase
  dynamics is flat, the Universe is static, and there are no bounce points. Such
  a situation is not interesting when searching for a bounce. For finite $\lambda$
  the qualitative description of the system does not depend on the specific value of
  this coupling constant, as it corresponds to the angle of the limiting cone
  and "shrinking" trajectories between double cone's branches. Therefore in order to simplify  further calculations we will set the value $\lambda=1$.

The bounce happens when a phase trajectory passes from the region $z<0$ to region
$z>0$, intersecting the plane $z=0$. At the crossing point $\dot
z>0$ must hold. Equation \ref{pe3} implies that this happens
if the crossing point is contained between lines $y=\sqrt{2}\,x$ and
$y=-\sqrt{2}\, x$ laying on the plane $z=0$. Those lines are the $z=0$ section of
an elliptic cone $\frac1{6}\left(x^2-\frac{y^2}2\right)-z^2=0$, whose interior consists of
trajectories with $\dot{z}>0$ (eq. (\ref{pe3})). The area outside this cone is
filled with trajectories along which $\dot{z}<0$.

To find the finite critical points we set all right-hand-sides of
equations (\ref{pe1}-\ref{pe3}) to zero. This
gives rise to the conditions
\be
x=y = z=  0.
\ee

Stability properties of this point are
determined by the eigenvalues of the Jacobian of the system
(\ref{pe1}-\ref{pe3}). More precisely, one has to linearize
transformed equations (\ref{pe1}-\ref{pe3}) at each point. Inserting
$\vec{x}=\vec{x}_0+\delta\vec{ x}$, where $\vec{x}=(x,y,z)$, and keeping terms
up to 1st order in $\delta\vec{x} $ leads to an evolution equation of the form
$\delta\dot{\vec{ x}}=A\delta\vec{x}$. Eigenvalues of $A$ describe stability
properties at the given point.

At the finite critical point $O=(0,0,0)$, the matrix $A$ has 2 purely imaginary
eigenvalues, which implies there are closed orbits in the $xy$-plane encircling the
$z$-axis, i.e.  point $O$ lays on a center line surrounded by closed orbits.   It is interesting to note, that in general case $\lambda\neq1$ eigenvalues of $A$ are following: $(\sqrt{\frac{{2}}{1-3\lambda}},-\sqrt{\frac{{2}}{1-3\lambda}},0)$. Hence for $\lambda>1/3$ point $O$ lays on a center line, for $\lambda<1/3$ it is a saddle. Nonetheless, the latter case is excluded form our considerations, as we explained in the introduction.

 To find critical points that occur at infinite values of the parameters we
rescale the infinite space  $(x,y,z)$ into a finite Poincar\'{e} sphere by means
of the coordinate change:
\bea
x &=& {X \over 1-r} , \label{poin1} \\
y &=& {Y \over 1-r},  \label{poin2} \\
z &=& {Z \over 1-r} , \label{poin3}
\eea
where
\bea
X&=&r\sin\theta \cos\varphi,\\
Y &=&r\sin\theta \sin\varphi,\\
Z&=& r \cos\theta ,\\
r^2&=& X^2 + Y^2 +Z^2.
\eea
We shall use both Cartesian coordinates $(X, Y, Z) $ and spherical ones: $(r, \theta,\varphi)$. 
We also  rescale the
time parameter  $t$ by defining new time parameter $T$ such that: $d{T} = dt/(1-r)$. In these
coordinates our phase space  is contained within a sphere of radius one
-- infinity corresponds to $r=1$.

This is a conformal transformation, hence the limiting cone for phase
trajectories is $ Z^2= \frac16( X^2+ Y^2)$; all physical
trajectories are contained within this cone.
Bounce points are located on the plane $Z=0$ within the region bounded by
lines $ Y=\sqrt{2}\,X$ and $Y=-\sqrt{2}\,
X$. The region containing trajectories with
$\dot{Z}>0$ (i.e. with $H$ increasing) is bounded by an elliptic cone
$\frac1{6}\left(X^2-\frac{Y^2}2\right)-Z^2=0$. 

 After Poincar\'e transformation, equations (\ref{pe1}-\ref{pe3}) take the following form, written in the spherical coordinates $(r, \theta,\varphi)$:
\bea
r'&=& \frac{ (r-1) r^2}{48} \cos\theta \left[82+14 \cos2 \theta-42 \cos2\phi\right.\nonumber\\
&&\left.+21 \cos2 (\theta-\phi)+21\cos 2(\theta+\phi)\right],\label{sph1}\\
\theta'&=& \frac{1}{24}r \sin\theta (5+7 \cos2\theta) (1+3 \cos2 \phi) ,\label{sph2}\\
\phi'&=& r-1-3r \cos\theta\cos\phi  \sin\phi.\label{sph3}
\eea
The form of the above equations is similar to the ones obtained in \cite{frolov,frolov2}.
Taking limit $r = 1$ and putting r.h.s. of equations for $\theta'$ and $\phi'$ to zero, we find 12 solutions
for $\theta,\phi$ at the Poincar\'e sphere,  shown in the Table 1.
As we can see, there are 4 saddle points
(more precisely saddle lines with end points at $S_1$,  $S_2$,  $S_3$,
$S_4$). In the contracting part of the phase portrait ($z<0$) there are two
attracting nodes $A_2$ and $A_4$ and two repulsing lines starting at $R_1$ and
$R_3$. Hyperbolic areas near the nodes are bounded by repulsing lines which play
role of separatrices. The expanding part is a mirror ("reversed in time") of the
contracting one. Stability properties of those point does not depend on the value of parameter $\lambda$, unless it is in the range $(1/3,\infty)$.

Stability properties of infinite critical points are described in the Table 1,
their position in 3D phase space, on a Poincar\'{e} sphere,  is shown in Fig. 1.
\begin{table*}[t]
\begin{center}
\begin{tabular}{|c|c|c|c|}
Point & $\varphi $ & $\theta$ &  Stability  \\
\hline
\hline
 S1&$\arcsin \sqrt{2/3}$ &$\pi/2$& Saddle line \\
\hline
 S2&$\pi-\arcsin \sqrt{2/3}$ &$\pi/2$& Saddle  line \\
\hline
 S3&$\pi+\arcsin \sqrt{2/3}$&$\pi/2$& Saddle  line \\
\hline
 S4&$2\pi-\arcsin\sqrt{2/3}$&$\pi/2$& Saddle line \\
\hline
 A1&0&$ \arccos \frac{\sqrt{7}}7$&Attracting line\\
 \hline
 R1&0&$ \pi-\arccos \frac{\sqrt{7}}7$& Repelling line\\
\hline
 R2&$\pi/2$&$ \arccos \frac{\sqrt{7}}7$&Repelling node\\
 \hline
 A2&$\pi/2$&$\pi- \arccos \frac{\sqrt{7}}7$& Attracting node\\
\hline
 A3&$\pi$&$ \arccos \frac{\sqrt{7}}7$&Attracting line\\
 \hline
 R3&$\pi$&$\pi- \arccos \frac{\sqrt{7}}7$& Repelling line\\
 \hline
 R4&$3\pi/2$&$ \arccos \frac{\sqrt{7}}7$&Repelling node\\
 \hline
 A4&$3\pi/2$&$\pi- \arccos \frac{\sqrt{7}}7$& Attracting  node\\
   \end{tabular}
\end{center}
\caption[crit]{\label{crit} The properties of the infinite critical points.}
\end{table*}

\begin{figure}
\begin{center}
\includegraphics[height=70mm]{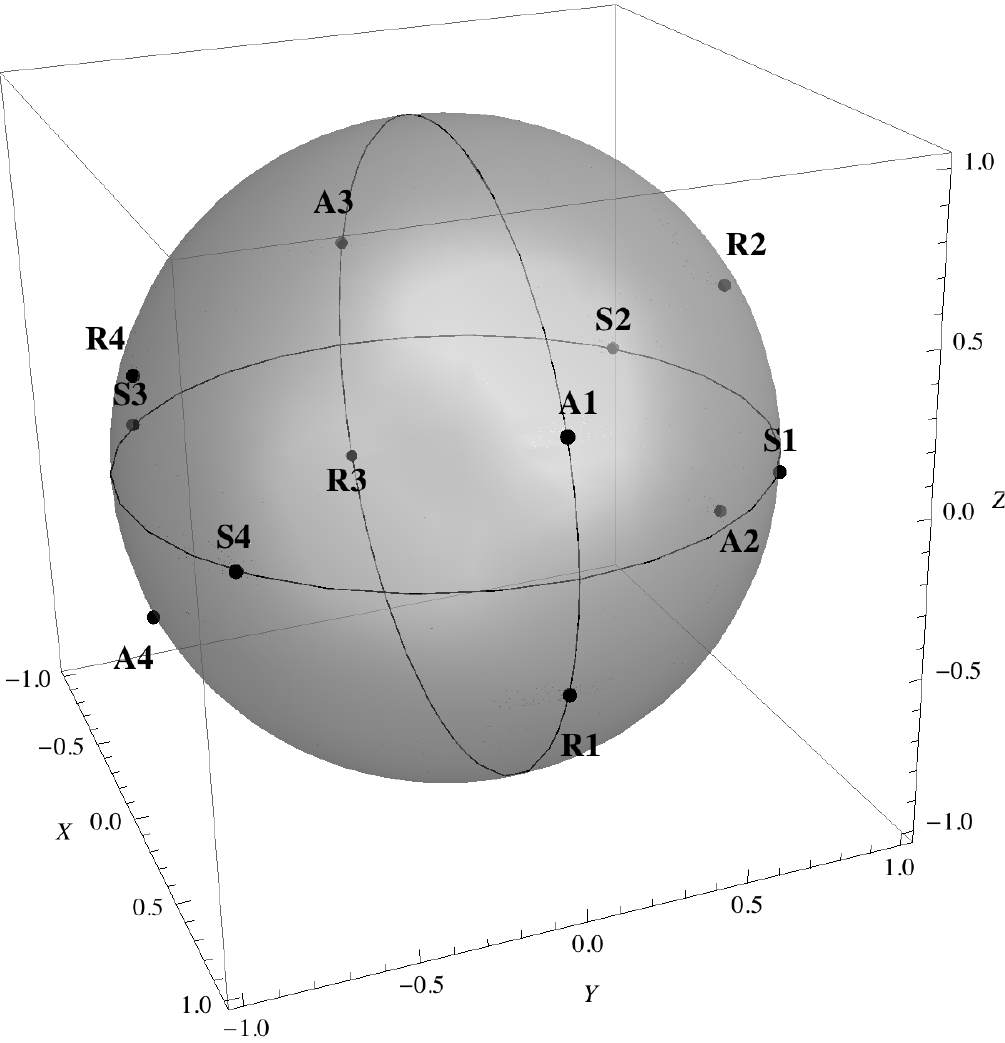}\label{first}
\caption[1]{Infinite critical points located on  a Poincar\'{e} sphere}
\end{center}
\end{figure}

\subsection{Trajectories}

When spatial curvature $k=0$ then  phase trajectories lay on the limiting cone $ \tilde{z}^2=
\frac16( {X}^2+ {Y}^2)$, as shown  in the Figure 2. In the contracting part all trajectories start winding around $z$-axis, then some of them end at attracting node $A_2$, some at $A_4$. Those two families are separated by repelling lines with end-points at $R_1$ or $R_3$, acting as separatrices. Expanding part is a mirror reflection with time reversed of the contracting part.
\begin{figure}
\begin{center}
\includegraphics[height=60mm]{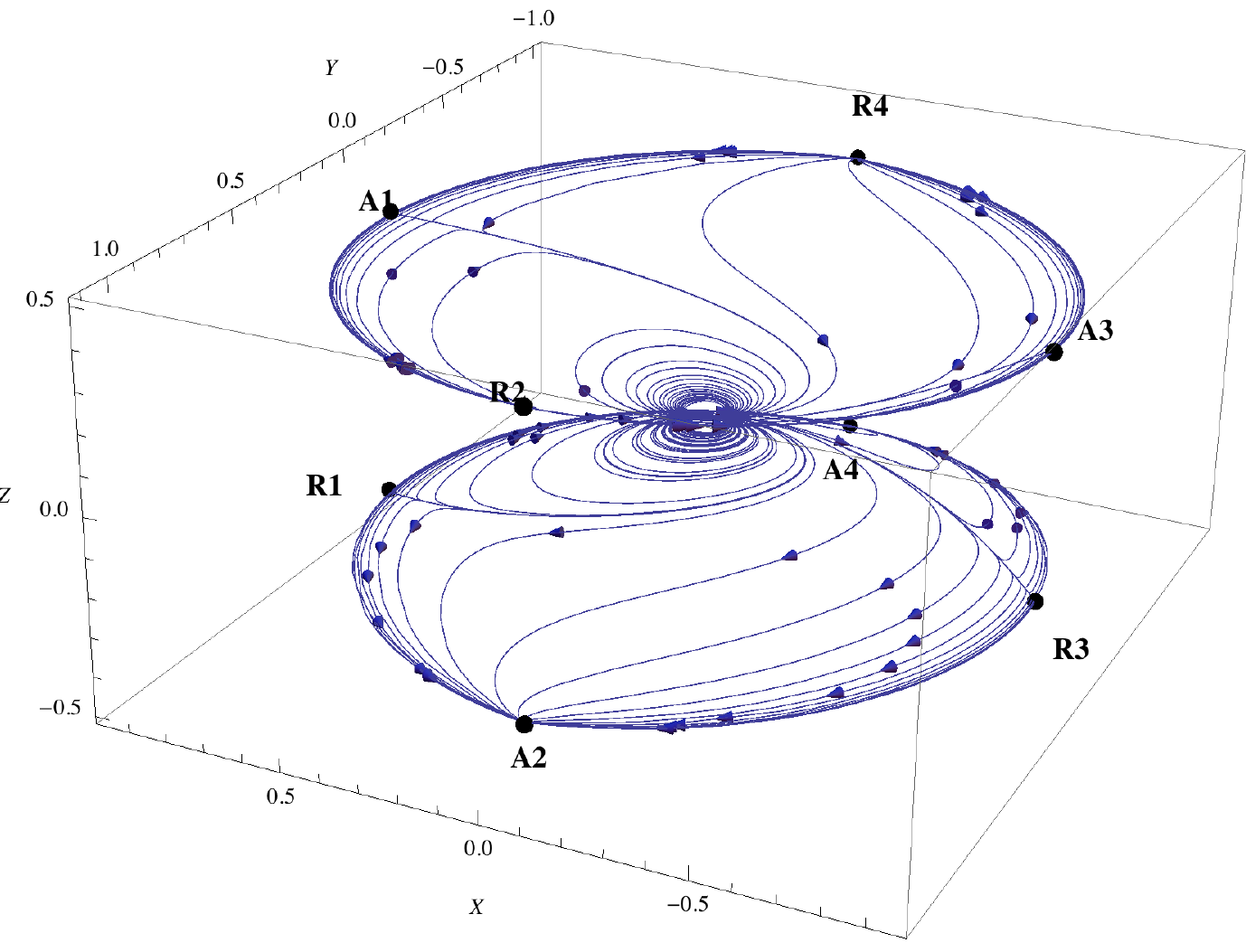}
\caption{Phase trajectories for flat HL universe}
\end{center}
\end{figure}

Trajectories of non-flat universes lay inside the limiting cone of flat space.
 In the
contracting part of the diagram ($Z<0$), trajectories start spiraling outside
from circles around the $Z$-axis. There are two families of such trajectories,
separated by repelling lines ending at $R_1$ and $R_3$.  In each family, there
are two possible scenarios for subsequent evolution. The first one, shown in
Figure 3a, is to end at an attractor node ($A_2$ or $A_4$), which also lays in
the contracting part of the phase diagram. On the way between $O$ and $A_2$ or
$A_4$, a trajectory may go up through $Z=0$ surface, undergoing a bounce there,
and then recollapse, crossing the $Z=0$ plane again, or go straight to the
attractor node, without bounce. In either case, the end is a Big Crunch.

The second scenario is shown on the Figure 3c. Here, after some oscillations and
$H$ decreasing, the trajectories reach an attractor -- a repelling line (that
ends either at $R_1$ or $R_3$), along which they move until $\dot{H}=0$.  Then
they rapidly go up, crossing the $Z=0$ (i.e. $H=0$) plane, undergoing a
bounce. After that, and after a period of accelerated expansion, they reach
another attractor -- an attracting line laying in the expanding part, with
endpoint at either $A_1$ or $A_3$. Along this line trajectories approach the
$Z$-axis, winding around it. A subcase of this scenario is shown on the Figure
3e, where trajectories do not go through accelerated contraction and expansion,
but cross the $Z=0$ surface during oscillations around the $Z$-axis.  This is in fact the scenario
described in \cite{Brandenberger:2009yt}.

Trajectories may also start at repelling nodes $R_2$ or $R_4$ in the expanding
part of the diagram. At those points $H=\infty$, i.e. there is a Bing Bang
there. After that and a period of extreme, but with decreasing rate, expansion,
there are again two possible scenarios. One is shown on the Figure 3b, where
trajectories reach an attracting line and end up winding around the
$Z$-axis. Before that, some of them collapse, crossing the $Z=0$ plane, then
slow down and finally stop contraction culminating in a bounce. Others show only
slow expansion, without crossing $Z=0$.

The last scenario is shown on the Figure 3d. Trajectories start at Big Bang
points $R_2$ or $R_4$, and after a period of slowing down expansion, reach a turning
point and start accelerated contraction, ending at Big Crunch points $A_2$ or
$A_4$.

For better visualization we have gathered some described families of trajectories Figure 3f.

\begin{figure}
\subfigure[][]{\includegraphics[height=45mm]{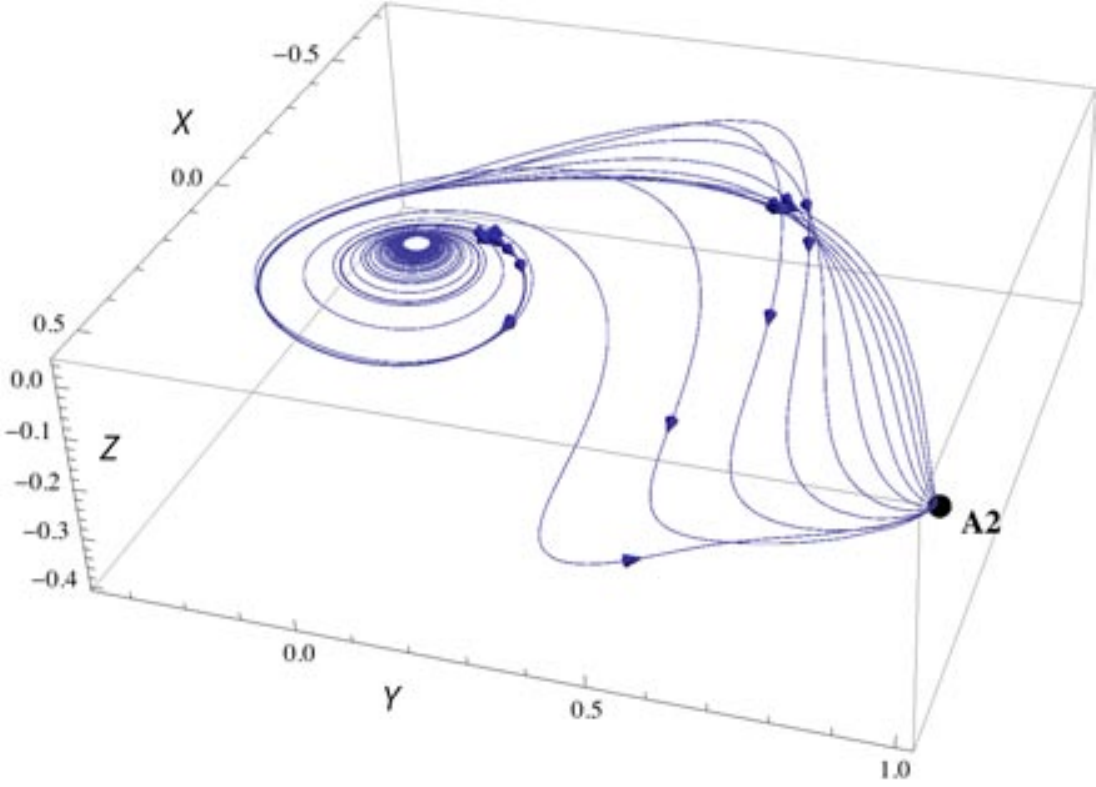} }
\subfigure[][]{\includegraphics[height=45mm]{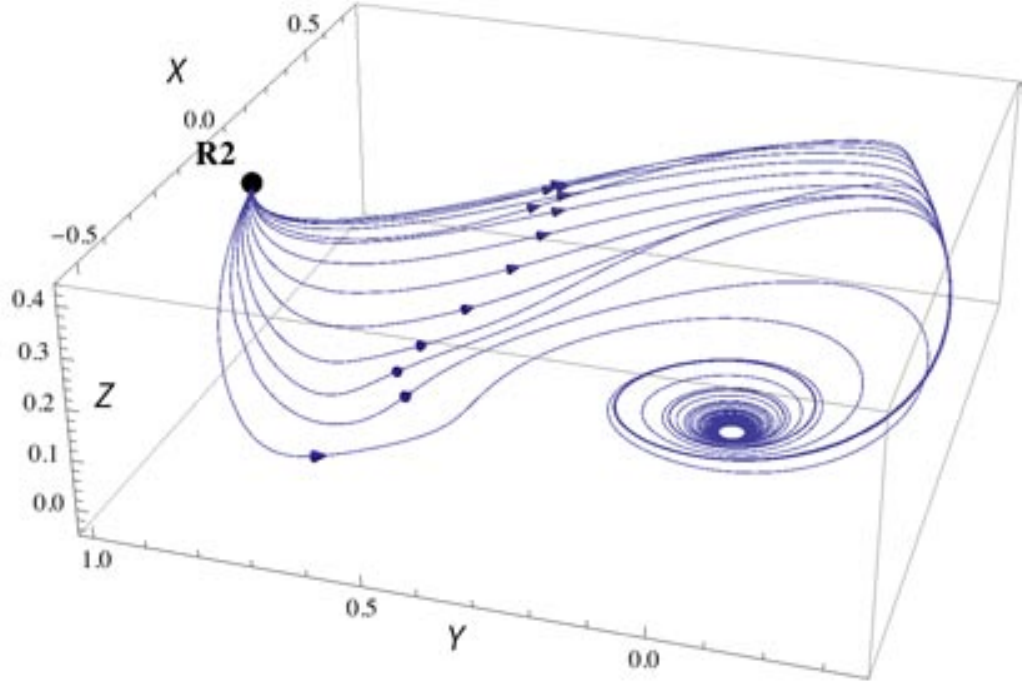} }
\subfigure[][]{\includegraphics[height=45mm]{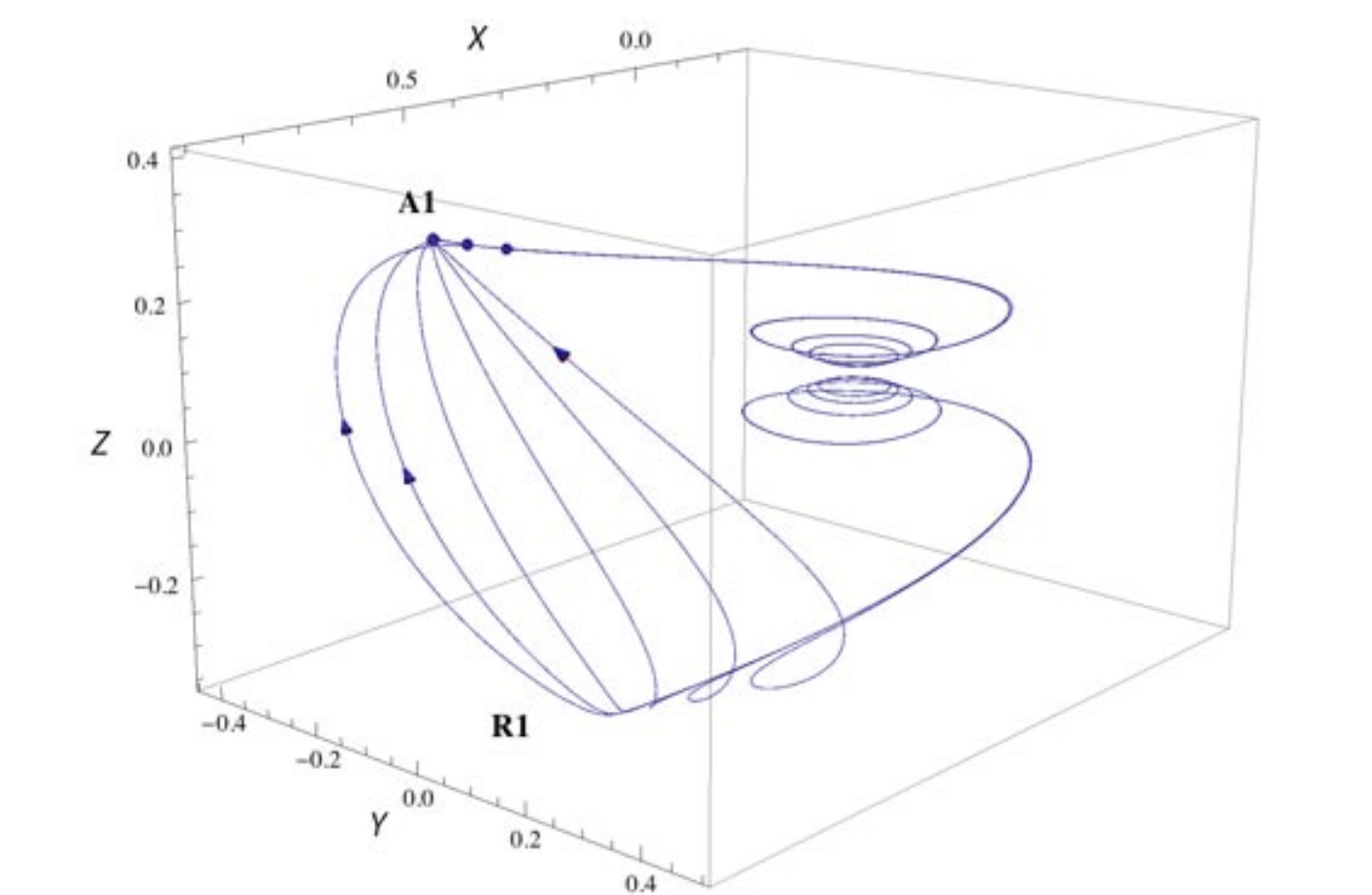} }
\subfigure[][]{\includegraphics[height=45mm]{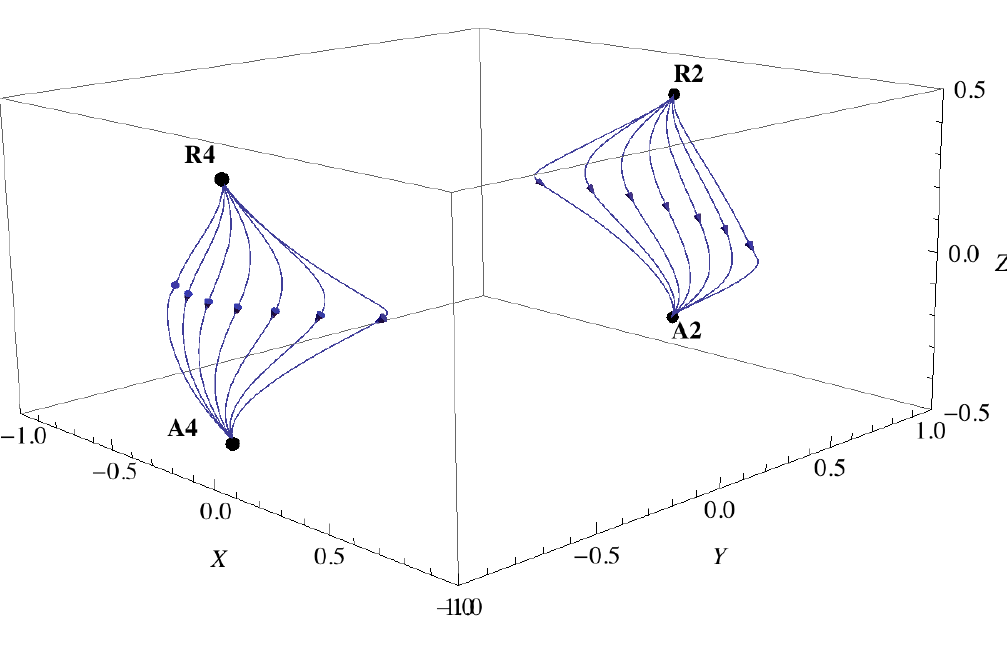} }
\subfigure[][]{\includegraphics[height=45mm]{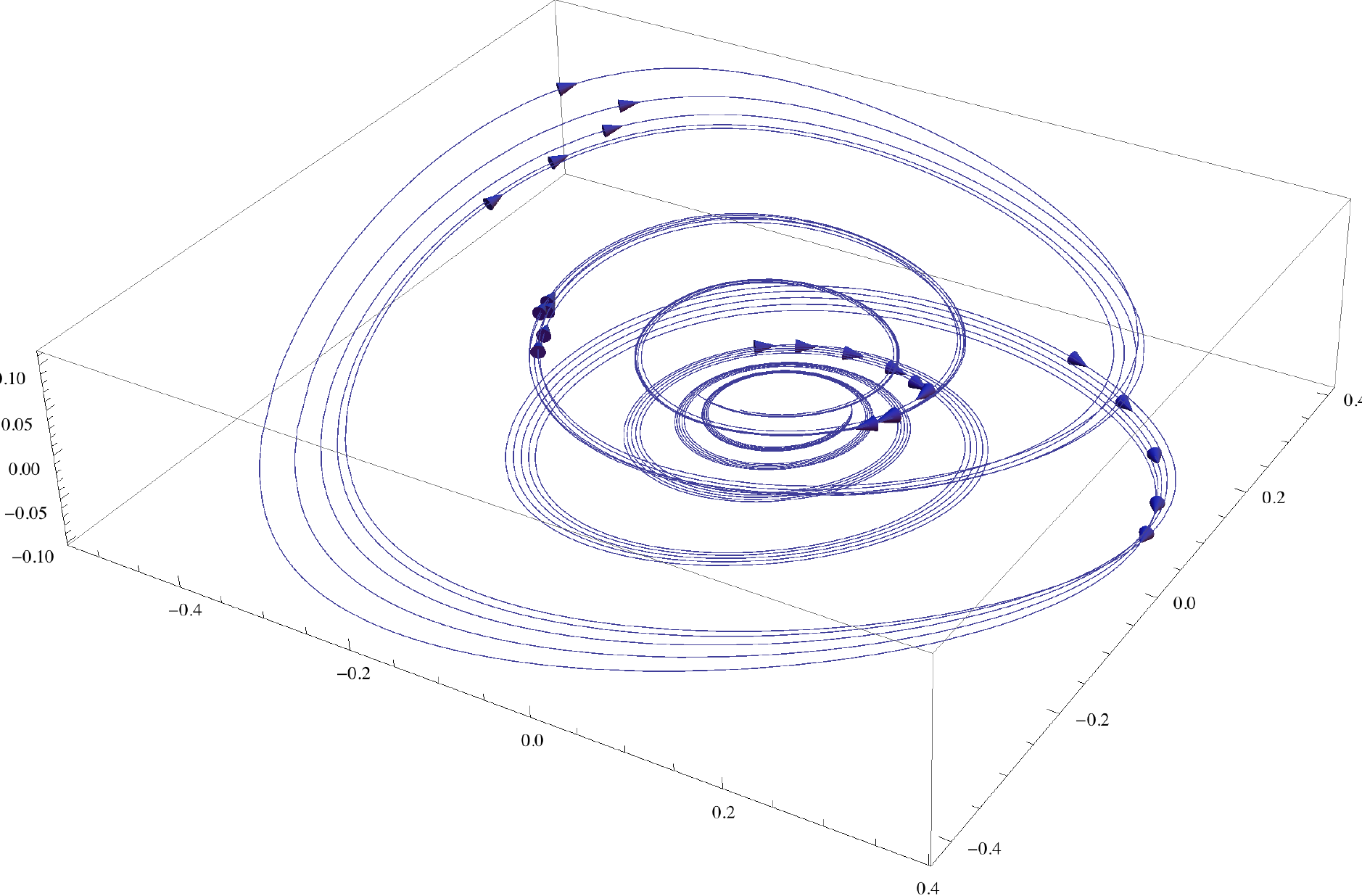} }
\subfigure[][]{\includegraphics[height=50mm]{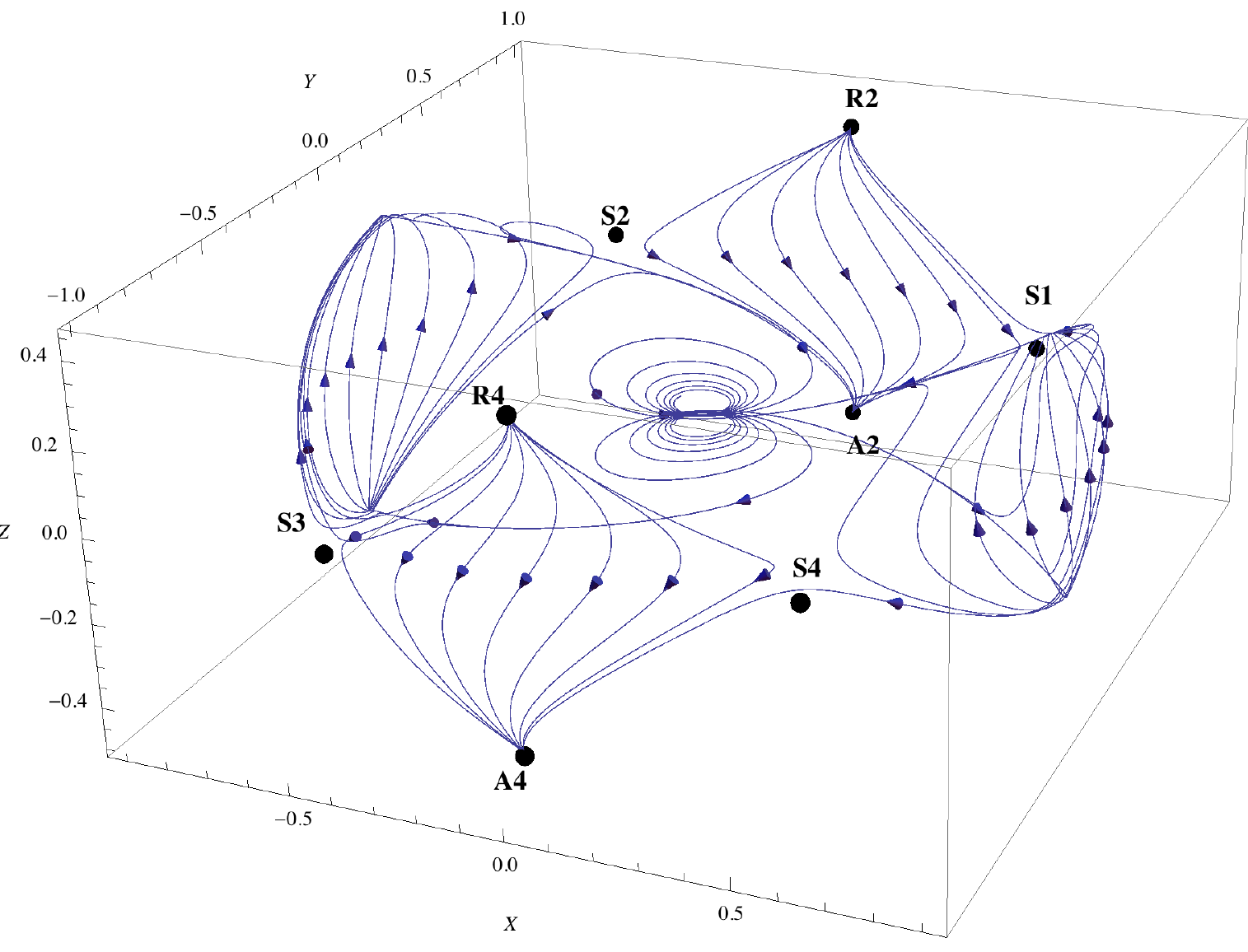}}
\caption{Different types of phase trajectories for a non-flat Ho\v{r}ava-Lifshitz universe}
\end{figure}

Special attention has to be paid to circular motion around $z$-axis. As stability properties of the point $(0,0,0$) and constraint equation (\ref{pe50}) suggest,  there may exist closed circular orbits laying on a  $Z=0$ plane ($H=0$). But they cannot. Equation (\ref{pe3}) does not allow for this, as $\dot{Z}=0$ is fulfilled only on a class of curves laying on $X^2-\frac{X^2}2-6Z^2=0$, i.e. on a surface of the elliptic cone mentioned before. Yet  numerical simulations
exhibit oscillating solutions, such as $X^2+Y^2=\mbox{\rm const.}$ and $Z$ oscillated around zero, $\dot{Z}>0$  between lines $Y=\sqrt{2}\,X$ and $Y=-\sqrt{2}\, X$,  $\dot{Z}<0$ outside this region. Such a trajectory resembles deformed circle. These solutions appear for sufficiently small $X$ and $Y$, for larger values of $X$ and $Y$ numerical simulations show slow decreasing of the radius of this "circle" . 

Finally, note that except for the special solution shown in Figure 3e -- which is the
bounce described by Brandenberger \cite{Brandenberger:2009yt}, there are also
other types of bounces. One, probably the most interesting, is shown in Figure
3c. Here a big existing universe slowly starts to contract, but later on the
contraction becomes exponential, until a bounce is reached and an exponential
expansion begins, which finally slows down.  Another type of bounce, shown on
the Figure 3a, happens again when a big universe slowly starts contracting,
stops and goes through an expanding phase for a while, then recollapses and ends
at Big Crunch.  The last one, shown on the Figure 3b, happens during a
transition from a Big Bang to a quasi stationary final stage (with $H$ slowly decreasing), however with a bounce on the way.

 Trajectories shown on the figures discussed above are numerical solutions
  of the equations (\ref{sph1}-\ref{sph3}). To find different bounce scenarios
  we investigated initial conditions: $\theta=\pi/2\pm0.01$,
  $\phi=i\frac{\pi}{20}$ ($i=1\ldots20$), each for $r=j/10$ ($j=1\ldots9$) and
  $r=0,9+j\cdot 0.01$; time in range $[-20,20]$. This procedure picked up the
  classes of trajectories discussed above. In general it may not be exhaustive
  in the sense that qualitatively different behavior of solutions may be
  possible. However it is sufficient for the purpose of understanding how
  bouncing scenarios emerge here due to the specific modification of general
  relativity which appears in Ho\v{r}ava's theory.

\section{Discussion and conclusions}

In this paper we have investigated the cosmological bounce in
Ho\v{r}ava-Lifshitz gravity. Using a 3D flow visualization technique we have
found that phase portraits in the considered theory have a different structure
than in standard cosmology. Comparing to results from the paper \cite{frolov2},
we can see that here are additional repellers ($R_1$ and $R_2$) in the
contracting part of a phase space, and mirror attractors in the expanding part.
Their presence allows the existence of a bounce, because now there are possible
new families of trajectories, starting at additional repellers in the
contracting part, and possibly ending at new attractors in expanding part, or
surrounding the $(0,0,0)$ point, which is now a center, compared to saddle in
standard cosmology. Those are realizations of the bounce. One of them is the
solution with oscillatory behavior described in (\cite{Brandenberger:2009yt});
there are however additional possibilities. The most interesting one contains a period of rapid contraction, and -- after a bounce -- a period of rapid expansion, what may fit inflationary scenario.

Nevertheless there are
still initial conditions which lead to a Big Crunch, as shown on the Figures 3a
and 3d, or which start at initial singularity (Fig. 3b and 3d). Hence the existence of
a bounce is not generic for Ho\v{r}ava theory and depends on initial
conditions.

Another interesting class of solutions consists of quasi stationary universes. These
solutions are described in phase space by closed  orbits, winding around the critical point
$(0,0,0)$ - a center.  All trajectories in the neighborhood of this point end up as closed orbits, "deformed circles".  Equations of motion do not allow for closed orbits laying on $Z=\mbox{const.}$ plane, resulting in slight deformation of the circular orbits. The values of $H$ oscillate  around stationary stage, for  sufficiently small values of $\varphi$ and $\dot{\varphi}$.  Values of scale parameter $a$ during this evolution are much bigger than the regime for which our simplifications are valid.
Therefore this behavior is not a feature of Ho\v{r}ava-Lifshitz theory,  but of cosmologies with
modified equations of motion, i.e. with the additional term $\sim 1/a^4$ in the Friedmann equations. Still, presence of this term leads to a different solution than
induced by  a negative potential as in \cite{frolov2}, due to different stability properties of finite critical points there.

 The visualizations described in this paper  describe the  dynamics of
Ho\v{r}ava-Lifshitz universe with vanishing cosmological constant $\Lambda$, or HL universe with non-zero $Lambda$ in the region of small scale factor $a$.
 Even in such slightly limited framework
they answer the question of possible scenarios realizing a bounce, and whether
it is generic for the theory or not. It appears not, as we have found solutions
leading to a Big Crunch, or starting at Big Bounce, both staying within the
regime of small $a$. There is also an interesting possibility of quasi stationary, oscillating
universe, existence of which is clearly implied by dark radiation term in
Friedmann equations.

Finally, it is worth stressing that the analysis presented here should be
applicable to other theories which lead to modifications of the Friedmann
equations.

\begin{center}
{\bf Acknowledgements}
\end{center}

I would like to thank Micha\l\ Spali\'nski for fruitful
discussions.

This work has been
supported by the Polish Ministry of Science and Higher Education
grant   PBZ/MNiSW/07/2006/37.

\end{document}